\documentclass[conference]{IEEEtran}
\IEEEoverridecommandlockouts
\usepackage{cite}
\usepackage{amsmath,amssymb,amsfonts}
\usepackage{graphicx}
\usepackage{textcomp}
\usepackage{xcolor}
\usepackage{array}
\usepackage{url}
\newcolumntype{L}[1]{>{\raggedright\arraybackslash}p{#1}}
\graphicspath{{figs/}}

\begin{document}

\title{A No-Code Toolchain for Validating Wearable\\Motion Sensing Systems
\thanks{% REVIEWER: add funding agency here, or delete this \thanks{}.
Identify applicable funding agency here.}
}

\author{\IEEEauthorblockN{Yao Zhang, Zhuchenyang Liu, Yu Xiao}
\IEEEauthorblockA{\textit{Department of Information and Communications Engineering} \\
\textit{Aalto University}\\
Espoo, Finland \\
yu.xiao@aalto.fi}
}

\maketitle

\begin{abstract}
Designing e-textiles for human motion sensing often demands engineering
expertise for sensor data processing, typically relying on advanced machine
learning methods. In addition, validating designs---collecting sensor
measurements and ground-truth motion data, training and testing machine
learning models, and iterating on sensor layouts---can be labour- and
time-intensive. To address this, we present a no-code toolchain that enables
non-technical users to validate system designs efficiently. The tool provides
three key capabilities: (1) deriving ground-truth joint-angle measurements from
webcam recordings; (2) training and testing machine learning models using
collected sensor data; and (3) visualising and analysing motion sensing
performance and sensor layout choices. Our tool requires a simple, low-cost
setup and produces test reports within minutes. A user study was conducted to validate the usability of this toolchain. This toolchain is open source and available at https://github.com/yaozhang182/etextile-validation-gui.
\end{abstract}


%==============================================================
\section{Introduction}
%==============================================================

E-textile garments (e.g., instrumented sleeves, knee bands, suits, and gloves)
that integrate strain, flex, and/or pressure sensors have been used to capture
upper- and lower-body motion \cite{tan2025knitted,galli2023knee,zhou2023mocapose}.
From an application perspective, the motion sensing performance must be
evaluated by comparing motion estimated from sensor measurements against
ground-truth kinematics. The garment design, including sensor design and layout
choice, should then be updated based on this evaluation. In other words,
validating a garment design requires a quantitative assessment of achievable
motion sensing performance. In practice, this process requires engineering
expertise and often access to specialised infrastructure for data collection and
processing. For instance, researchers may use optical motion-capture systems,
either commercial solutions or open-source packages such as
Pose2Sim \cite{pagnon2022pose2sim}, to calculate kinematics as ground truth.
These systems require multiple high-speed cameras and careful setup and
calibration, making them costly and time-consuming. Moreover, the pipeline for
converting sensor readings into joint-angle estimates or other meaningful
information is typically released as bespoke research
code \cite{zhu2024cnnlstm,chhoeum2021capacitive,gholami2019running,tavassolian2020inductive},
which requires significant effort to adapt and reuse. Consequently, rapid design
iteration is challenging, particularly for individual designers or small teams
without an engineering background or access to specialised infrastructure.

The aim of this work is to develop a no-code toolchain that allows e-textiles
researchers and practitioners with no background in machine learning or computer
vision to validate their designs using a simple, low-cost setup and obtain
performance feedback within minutes. Specifically, the toolchain will (1) derive
ground-truth motion data (e.g., joint-angle measurements) from footage captured
with accessible hardware such as a webcam, eliminating the need for expensive
motion-capture systems; (2) automate the training and testing of machine
learning models using the collected sensor data and ground-truth motion data;
and (3) provide visualisation and analysis of motion sensing performance to
guide iteration on sensor selection and layout choices.

Individual tools already exist that partially address parts of this workflow.
For example, single-camera pose estimators such as
MediaPipe \cite{lugaresi2019mediapipe,bazarevsky2020blazepose} can provide
joint-position estimates with per-joint errors of only a few degrees, and have
been benchmarked against marker-based systems across several kinematic
tasks \cite{lafayette2023validation}. Although their errors exceed those of
marker-based systems and fall short of clinical gait-analysis requirements,
prior work shows that MediaPipe can track joint kinematics with sufficient
consistency for comparative analysis and iterative
evaluation \cite{lafayette2023validation}. Lightweight regression models for
joint-angle estimation, implemented with libraries such as Scikit-learn or
PyTorch, can be trained on a standard laptop within
minutes \cite{zhu2024cnnlstm,chhoeum2021capacitive}. However, these components
remain siloed and have not been integrated into an end-to-end, no-code workflow
accessible to researchers and practitioners with no background in machine
learning or computer vision.

We address this gap by developing an end-to-end toolchain as a desktop
application with a graphical user interface (GUI). The design and implementation
of the toolchain are presented in the following sections.We also report a usability study that evaluates how well the toolchain can be used by users.

%==============================================================
\section{System Architecture and Workflow}
\label{sec:architecture}
%==============================================================

\begin{figure}[t]
  \centering
  \includegraphics[width=0.72\columnwidth]{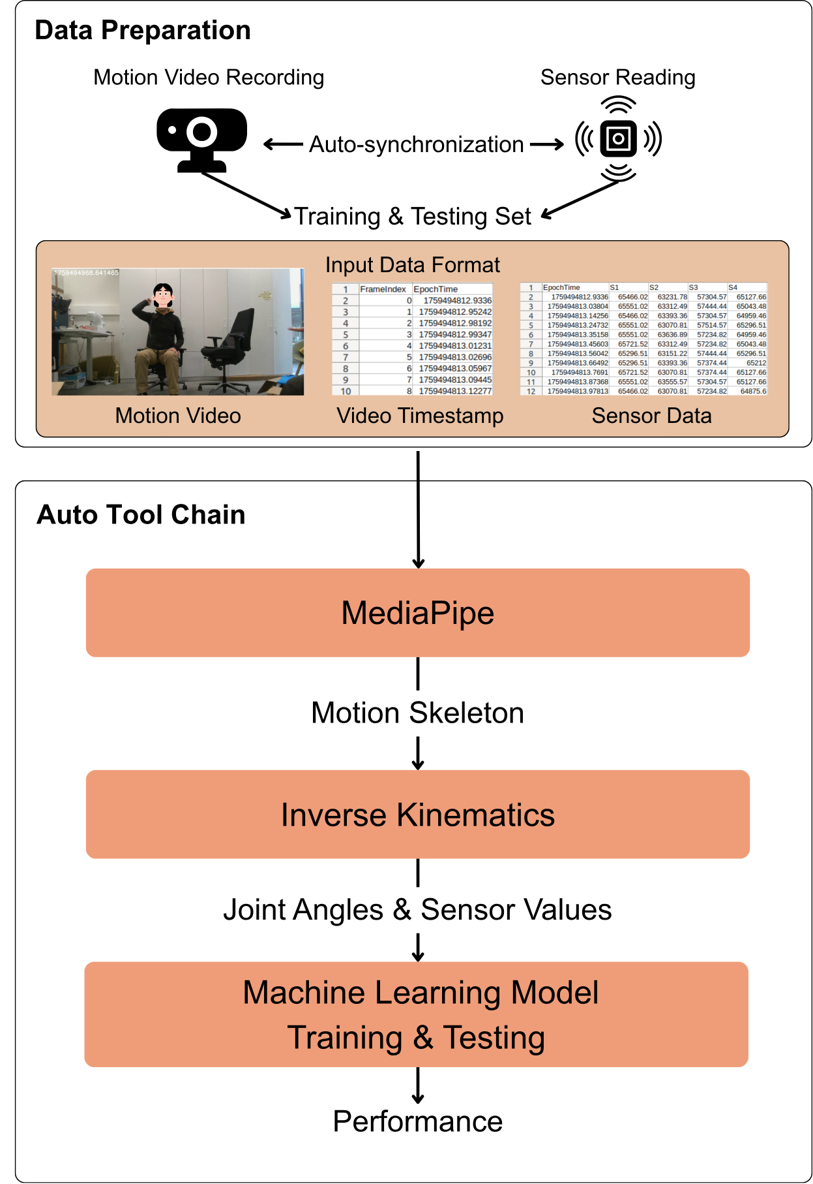}
  \caption{Overview of the toolchain pipeline. Two time-synchronised inputs
  (webcam video and sensor CSV) are processed by MediaPipe Pose, an
  inverse-kinematics module, and a regression model to produce per-joint motion
  sensing performance.}
  \label{fig:pipeline}
\end{figure}

As illustrated in Fig.~\ref{fig:pipeline}, the toolchain takes two synchronised
inputs: a webcam recording of a subject who performs movements while wearing the
garment under test, and the sensor data recorded by the garment at the same
time. Section~\ref{sec:data} describes the input format in detail.

The input is processed in four steps. First, per-frame 3D body landmarks are
extracted from each video with MediaPipe Pose \cite{lugaresi2019mediapipe}.
Second, an inverse-kinematics module converts the landmark sequence into
joint-angle channels following standard biomechanical conventions~\cite{wu2002isb1,wu2005isb2}, and these
are aligned with the sensor data on the shared clock. Third, the user selects
sensor channels as inputs and joint-angle channels as targets, and a regression
model is trained to map the former to the latter. Fourth, the trained model is
evaluated on sessions held out from training, and per-joint performance is
reported.

The full pipeline runs on a standard laptop CPU and requires no motion-capture
system or optical markers.

%==============================================================
\section{Graphical Interface and Implementation}
\label{sec:gui}
%==============================================================

All steps of Section~\ref{sec:architecture} are carried out through a desktop
application with five tabs, arranged in the typical order of use
(Figs.~\ref{fig:tab1}--\ref{fig:tab5}). Users only select options and press
buttons. No code is written and no external script is run. This section first
describes the input data and then each tab together with the part of the
pipeline it runs. The application is written in Python~3.10 with PyQt6 for the
interface, MediaPipe~0.10 for pose estimation, PyTorch for the regression
models, and pandas and NumPy for data handling.

% Every tab remains accessible
% at all times, and a tick mark in the tab bar shows which steps have been
% completed. Intermediate results---skeleton sequences, joint-angle channels, the
% trained model and the evaluation metrics---are kept in memory, so the user can
% return to an earlier tab without repeating later work. A \emph{Reset} button in
% the tab bar clears the current analysis, after confirmation, and returns to the
% first tab.

% Because the intended users are not in a position to audit the underlying models,
% the application also reports how reliable its ground truth is. MediaPipe's
% per-landmark confidence is propagated to every joint-angle channel and shown at
% three points in the workflow: when the tracking is inspected
% (Section~\ref{sec:tab2}), when the regression targets are chosen
% (Section~\ref{sec:tab3}), and next to each error in the results
% (Section~\ref{sec:tab5}).

%--------------------------------------------------------------
\subsection{Input Data}
\label{sec:data}
%--------------------------------------------------------------

\begin{figure*}[t]
  \centering
  \includegraphics[width=0.85\textwidth]{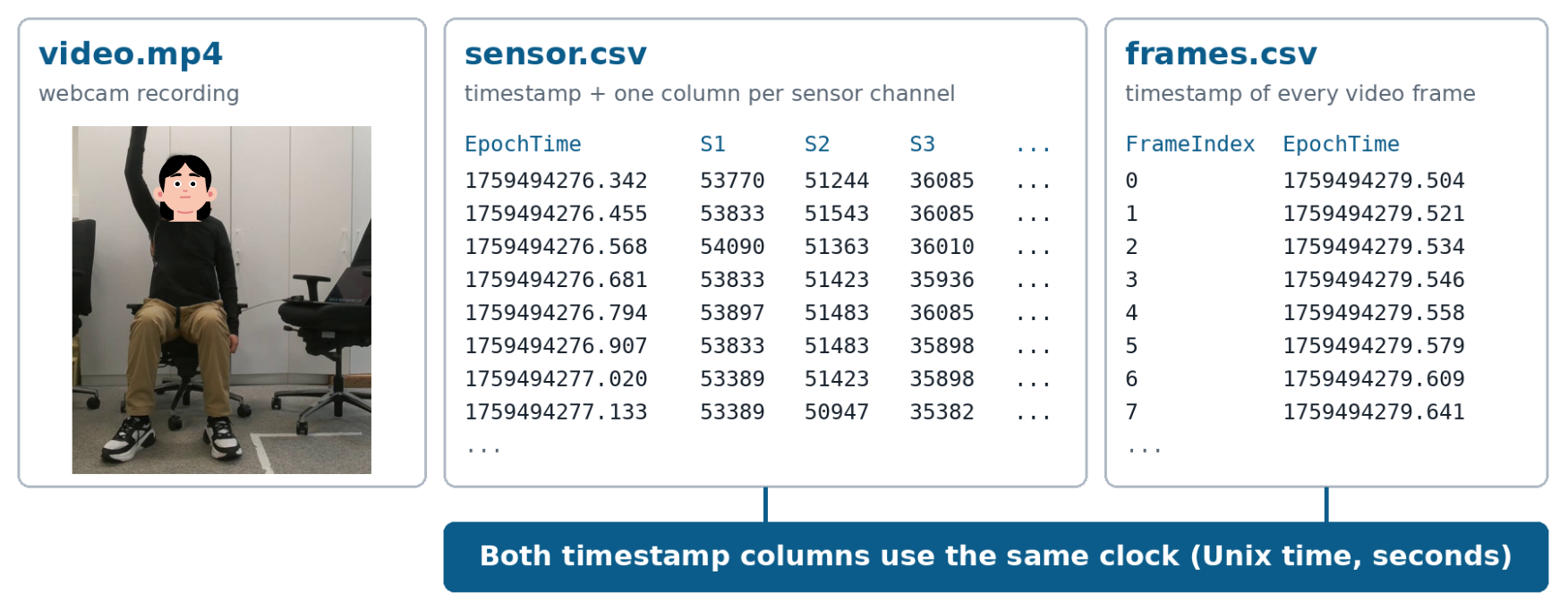}
  \caption{The three files of one session, shown for an example recording. The
  video is recorded with an ordinary webcam. The sensor CSV holds one timestamp
  and one column per sensor channel for each sample, and the frame-timestamp CSV
  holds the timestamp of every video frame. Both timestamp columns use the same
  clock, which lets the toolchain align the two streams.}
  \label{fig:data}
\end{figure*}

The toolchain organises its input by session. A session is one recording of one
wearer and consists of three files (Fig.~\ref{fig:data}). The first is the
video, recorded with a single webcam placed 1.5--3\,m in front of the subject
with the full body in frame. The second is the sensor CSV file, which contains an EpochTime column with the timestamp of each sample and one column for each sensor on the garment. The third is a frame-timestamp CSV file that gives the timestamp of every video frame through its FrameIndex and EpochTime columns. Both CSVs
may be separated by commas or semicolons, and the separator is detected
automatically.

Both timestamp columns use the same wall clock (Unix epoch time), which is what allows the two streams to be aligned. The two devices may record at different rates. In the example recordings, the sensors are sampled at about 9,Hz and the video at about 30 frames per second. The two recordings also do not need to start or stop at the same moment, as this is difficult to control during real-world data collection. To address this, our software automatically performs time alignment and restricts training and testing to the overlapping interval covered by both streams. We also provide data-collection scripts, together with a set of example recordings, on our GitHub page.

% Two further terms are used in the rest of the paper. A sensor channel is the time
% series of a single sensor on the garment, stored as one column of the sensor
% CSV. A joint-angle channel is the time series of a single anatomical degree of
% freedom, such as right-shoulder flexion.

%--------------------------------------------------------------
\subsection{Data Import: Sessions, Validation and Pose Extraction}
\label{sec:tab1}
%--------------------------------------------------------------

\begin{figure*}[t]
  \centering
  \includegraphics[width=0.78\textwidth]{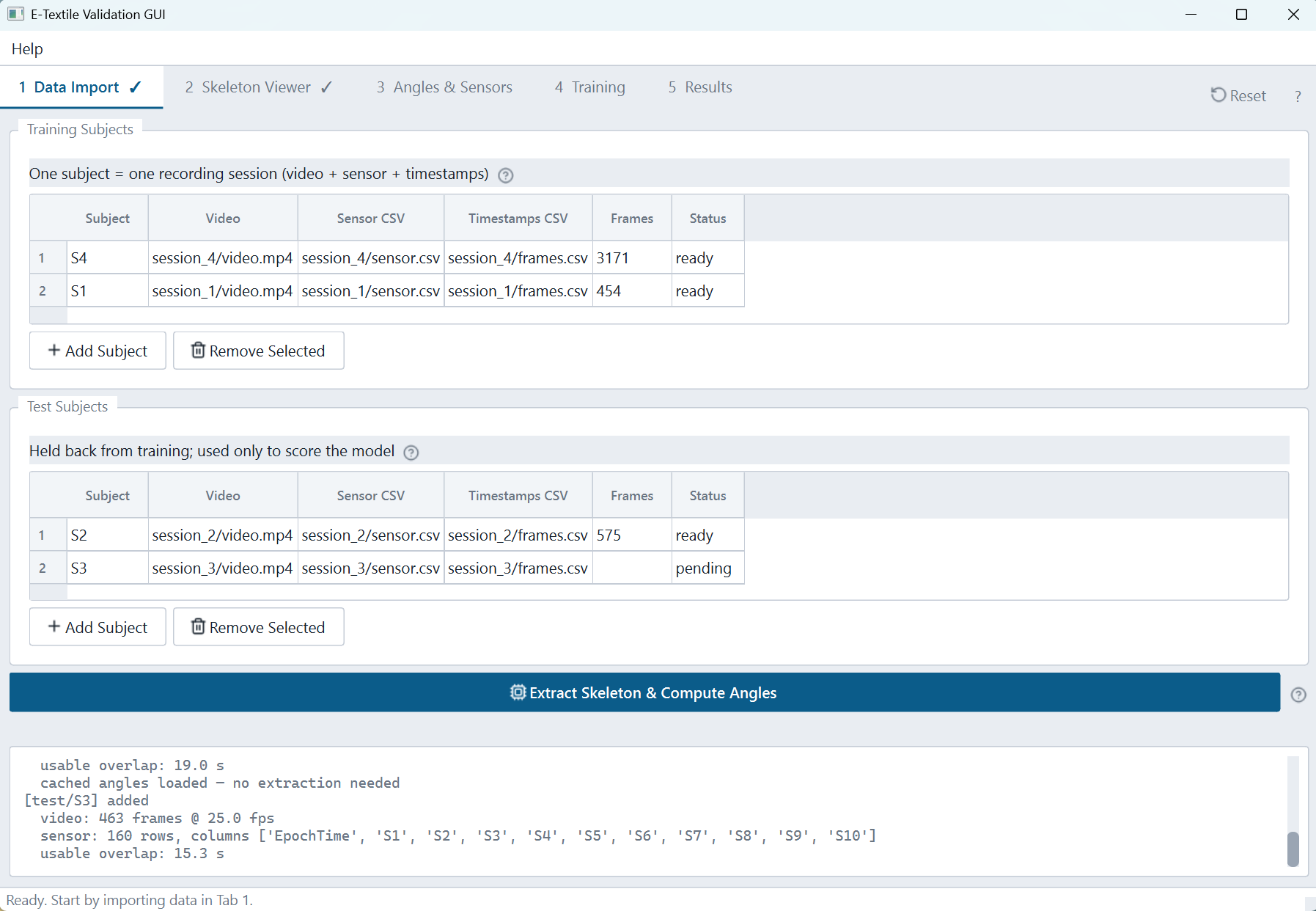}
  \caption{Data Import tab. Sessions are listed in separate training and test
  tables with their three files, frame count and processing status. The single
  extraction button processes every pending session. The log below reports, for
  each session, the frame rate derived from the timestamps, the detected sensor
  channels and the time window covered by both streams.}
  \label{fig:tab1}
\end{figure*}

The Data Import tab (Fig.~\ref{fig:tab1}) organises the input by session.
Sessions are added independently to a training table, whose sessions are used to
train the model, and to a test table, whose sessions are used to evaluate the
garment and the model. Validating a garment on several wearers is the normal
case, so any number of sessions can be added to either table. The Extract
Skeleton \& Compute Angles button then processes every session in a background
thread. Results are cached in the data folder, so a recording is processed once
and re-importing it later skips extraction.

Once the data are imported, the pose is extracted from every video frame with
MediaPipe \cite{lugaresi2019mediapipe,bazarevsky2020blazepose}, which returns 33
landmarks in metric 3D coordinates together with a visibility score for each
landmark. The landmarks are mapped onto the 22 body joints of the SMPL body model
\cite{loper2015smpl,romero2017mano} with a transform from MediaPipe's coordinate
convention into a right-handed body frame. MediaPipe has no pelvis or spine
landmarks, so the pelvis is placed at the hip mid-point and the spine joints are
interpolated between the hips and the shoulders. From the SMPL joint positions
the pipeline computes 26 joint-angle channels covering the pelvis, lumbar spine,
hips, knees, ankles, shoulders, elbows and neck, following the ISB
recommendations \cite{wu2002isb1,wu2005isb2}. Twelve of these channels can jump
between $+180^\circ$ and $-180^\circ$. They are unwrapped frame by frame so that
such jumps do not dominate the training loss.

%--------------------------------------------------------------
\subsection{Skeleton Viewer: Checking the Ground Truth}
\label{sec:tab2}
%--------------------------------------------------------------

\begin{figure*}[t]
  \centering
  \includegraphics[width=0.78\textwidth]{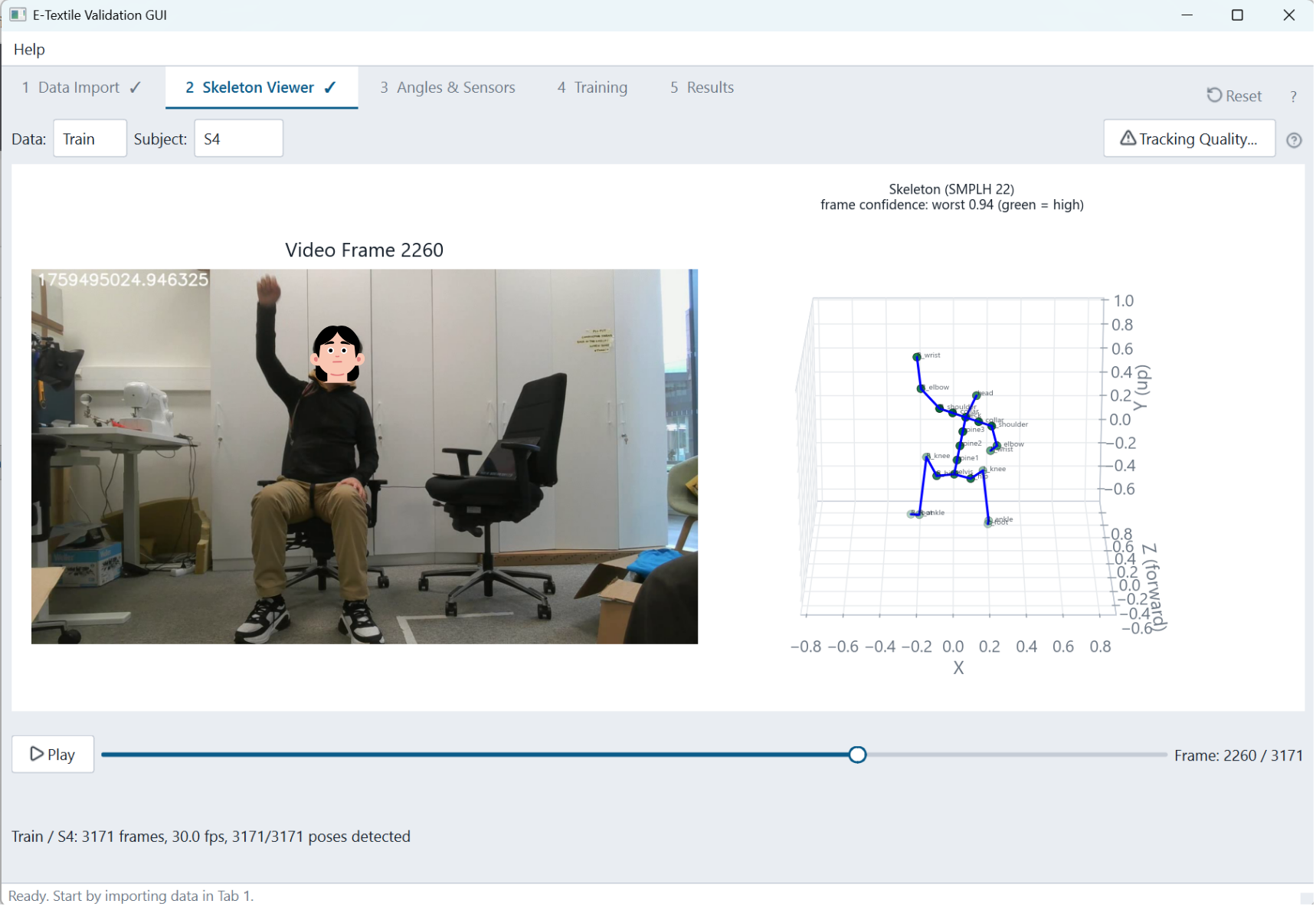}
  \caption{Skeleton Viewer tab. The video frame (left) and the reconstructed 3D
  skeleton (right) follow a shared timeline. Joints are coloured by tracking
  confidence on each frame, and the Tracking Quality button opens a summary of
  the pose estimation quality for the whole video.}
  \label{fig:tab2}
\end{figure*}

Every later result is only as good as the pose tracking, so the second tab lets
the user check it before training anything (Fig.~\ref{fig:tab2}). The original
video frame and the reconstructed 3D skeleton are shown side by side on a
scrubbable timeline with playback, and a train/test and session selector chooses
the recording.

Each joint in the skeleton is coloured on a green-to-red scale by its tracking
confidence from MediaPipe \cite{lugaresi2019mediapipe} on the current frame. A
user scrubbing through the recording therefore sees directly when and where
tracking degrades, for example when an arm passes in front of the body. The
Tracking Quality window summarises the whole recording with an overall rating,
the least reliable landmarks and joint-angle channels, and the fraction of frames
in which a landmark falls below 0.5, the level at which MediaPipe treats a
landmark as likely occluded.

% \emph{Confidence propagation.} MediaPipe's visibility score is the output of a
% classifier trained to flag landmarks that are occluded or poorly localised
% \cite{bazarevsky2020blazepose}. The 33 landmark scores are mapped onto the 22
% joints, and each joint-angle channel receives, frame by frame, the \emph{minimum}
% confidence of the joints it is computed from: an angle is only as trustworthy as
% the worst landmark feeding it. Channels that depend on the synthesised pelvis
% and spine are additionally marked as reconstructed rather than measured,
% independently of their visibility. The scores are ratings of tracking
% reliability, not error estimates, and the interface never converts them into
% degrees.

%--------------------------------------------------------------
\subsection{Angles and Sensors: Defining the Learning Problem}
\label{sec:tab3}
%--------------------------------------------------------------

\begin{figure*}[t]
  \centering
  \includegraphics[width=0.78\textwidth]{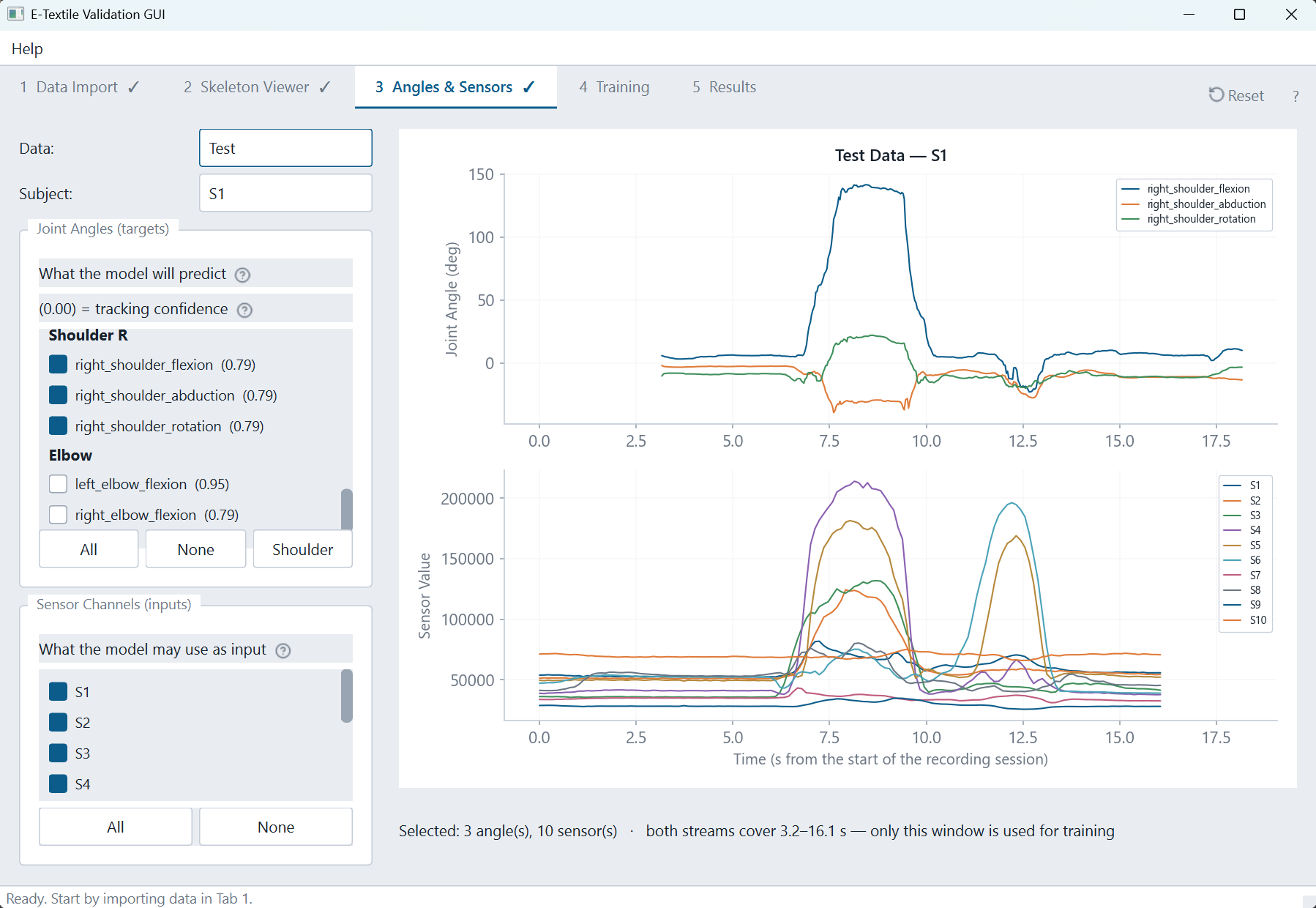}
  \caption{Angles and Sensors tab. The left panel lists joint-angle channels
  (targets, grouped by body part, each labelled with its mean tracking
  confidence) and sensor channels (inputs). The right panel overlays the
  selection on one time axis, and the line below the plots states the time
  window covered by both streams.}
  \label{fig:tab3}
\end{figure*}

The third tab is where the user decides what the model will predict and from
which sensors (Fig.~\ref{fig:tab3}). Joint-angle channels are listed as
checkboxes grouped by body part, with quick-selection buttons, and each is
labelled with its mean tracking confidence. Sensor channels are detected from
the CSV columns, and only channels present in every session are offered, which
prevents a selection from referring to a channel that some session lacks. The
selected angles and sensor channels are plotted in two panels that share one
time axis. Comparing them by eye is the quickest check of whether a sensor
responds to a movement at all, so both panels are placed on the same clock.

%  Two alignment modes are offered. The default interpolates the
% sensor stream linearly onto the video frame timestamps, which keeps every
% ground-truth frame; the alternative interpolates the joint angles onto the
% sensor timestamps.

%--------------------------------------------------------------
\subsection{Training: Models and Session-Aware Windowing}
\label{sec:tab4}
%--------------------------------------------------------------

\begin{figure*}[t]
  \centering
  \includegraphics[width=0.78\textwidth]{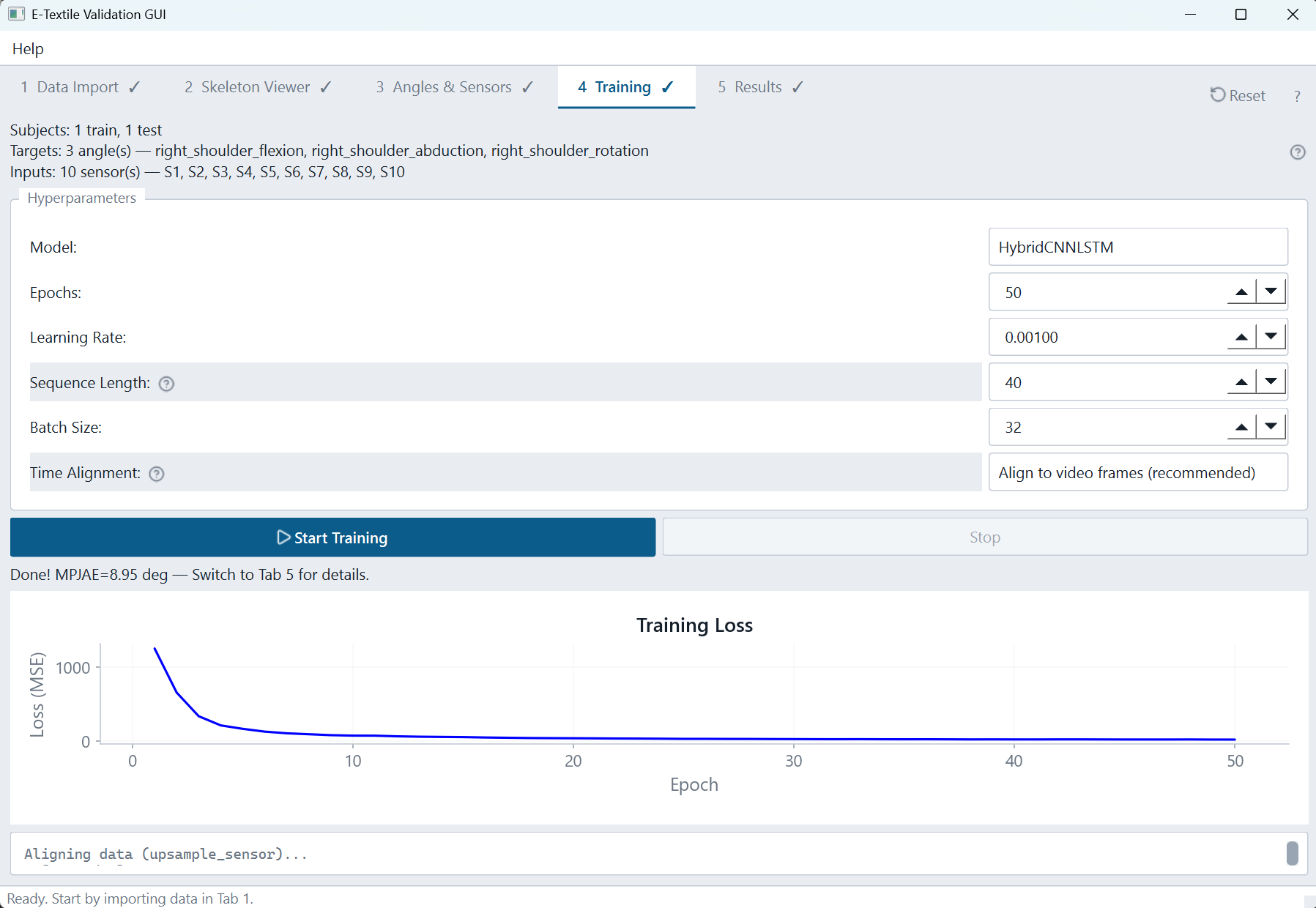}
  \caption{Training tab. A summary of the selected sessions, targets and inputs
  is shown above the hyperparameters, all of which have working defaults. The
  loss curve is updated after each epoch.}
  \label{fig:tab4}
\end{figure*}

The Training tab (Fig.~\ref{fig:tab4}) summarises what will be trained, namely
the number of training and test sessions, the target angles and the input
channels. It also lets the user adjust the hyperparameters: the model type,
number of epochs, learning rate, sequence length, batch size and alignment mode.
Two alignment modes are offered. The default interpolates the sensor stream
linearly onto the video frame timestamps, which keeps every ground-truth frame.
The alternative interpolates the joint angles onto the sensor timestamps.

Each aligned session is then cut into overlapping windows of $L$ consecutive
samples (default sequence length $L{=}40$), and each window is labelled with the
joint angles at its last sample. Windows are formed within each session and the
per-session sets are concatenated afterwards. Sensor values are standardised with
statistics fitted once on all training sessions and applied to the test sessions.

The default model is a hybrid CNN-LSTM. Two 1-D convolution layers (32 and 64
filters, kernel size 3, each followed by batch normalisation, ReLU and dropout of
0.1) extract features across sensor channels at every time step, a single-layer
LSTM with 64 hidden units models the temporal context, and a two-layer fully
connected head outputs one value per target angle. A lighter convolution-only
model (SimpleCNN1D) with global average pooling is provided as an alternative.
Both are trained on the CPU with the Adam optimiser and a mean-squared-error loss,
with the learning rate reduced when the loss plateaus and gradients clipped to
unit norm. The defaults are 50 epochs, a learning rate of $10^{-3}$ and a batch
size of 32.

%--------------------------------------------------------------
\subsection{Results: Metrics, Predictions and Sensor Importance}
\label{sec:tab5}
%--------------------------------------------------------------

\begin{figure*}[t]
  \centering
  \includegraphics[width=0.78\textwidth]{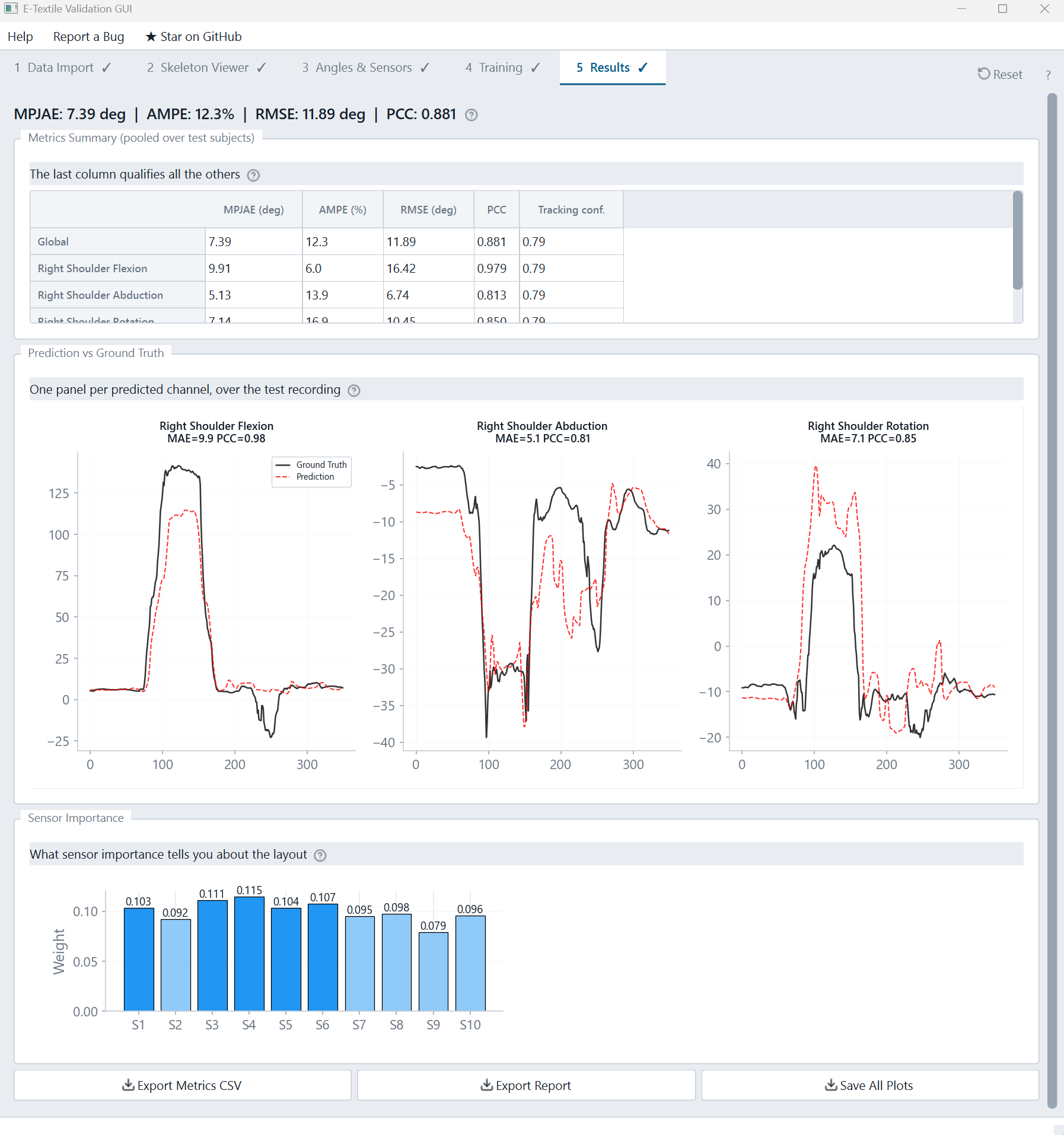}
  \caption{Results tab. The headline gives the four pooled metrics. The table
  reports them per target angle, with the tracking confidence of that angle's
  ground truth in the last column. Below are the predicted and ground-truth
  traces for each target, and the sensor-importance chart follows further down
  the tab.}
  \label{fig:tab5}
\end{figure*}

The last tab reports how well the selected sensors predict the selected angles
(Fig.~\ref{fig:tab5}). For a target angle with ground truth $y_t$ and prediction
$\hat{y}_t$ over $N$ test samples, the tab reports the mean per-joint absolute
error $\mathrm{MPJAE}=\frac{1}{N}\sum_t|\hat{y}_t-y_t|$, the root-mean-square
error (RMSE), the Pearson correlation coefficient (PCC), and the absolute mean
percentage error (AMPE). AMPE expresses MPJAE as a percentage of the angle's
range of motion in the test data,
$\mathrm{AMPE}=100\cdot\mathrm{MPJAE}/\max(y_{\max}-y_{\min},\,5^\circ)$, where the
$5^\circ$ floor prevents a nearly static angle from producing an inflated
percentage. Errors in degrees show how far off the prediction is, while PCC shows
whether it follows the shape of the movement. Below the metric tables, the
Prediction vs Ground Truth panel plots the predicted joint angle against the
ground truth over the test session, which shows where the error arises.

The sensor-importance chart supports layout decisions. For both models, the
importance of a sensor channel is the L1 norm of the first convolution layer's
weights connected to it, normalised over all channels. It indicates how much the
trained model draws on each channel, and a channel with consistently low
importance is a candidate for relocation or removal in the next garment
revision. The measure describes the trained model and makes no causal claim, so
it is best used to guide the next sensor selection, which the user can then test
by retraining. Metrics, a text report and all figures can be exported.

% When there is more than one test session, a per-session table is shown before the
% pooled one, so that a result driven by one unusual wearer is visible rather than
% averaged away. The pooled table has a final column with the mean tracking
% confidence of each target's ground truth. This column changes how the other
% columns should be read: a large error on a poorly tracked angle says little
% about the garment, because the reference itself was unreliable, whereas a large
% error on a well-tracked angle is a genuine finding.

%--------------------------------------------------------------
\subsection{Distribution, In-Application Help and Tutorial Video}
\label{sec:distribution}
%--------------------------------------------------------------

A no-code tool should also be easy to install. The application is therefore
distributed as a self-contained Windows bundle, built with PyInstaller by a
continuous-integration job on a Windows runner. The user unpacks the bundle and
starts it directly, with no Python installation, package manager or terminal
involved. For users who want to modify the pipeline, a setup script for
installation from source is available on GitHub. We also provide a user manual
with annotated screenshots that describes the input format, every tab and the
meaning of each metric. It is opened from the Help menu, and question-mark
buttons placed next to the less obvious controls explain that control and link
to the relevant section of the manual. In addition, the GitHub page offers a
tutorial video that demonstrates how to use the toolchain.

%==============================================================
\section{Technical Evaluation}
\label{sec:benchmarks}
%==============================================================

This section reports how long a validation cycle takes and summarises what the
literature says about the accuracy of the pose estimation the toolchain relies
on.

%--------------------------------------------------------------
\subsection{Processing Time}
\label{sec:timing}
%--------------------------------------------------------------

Table~\ref{tab:timing} breaks one validation cycle into its stages. It was
measured with the benchmark script included in the repository on the example
recordings, using three training sessions and one test session (4{,}663 video
frames, 155\,s of video in total). To approximate a laptop, the process was
restricted to four cores of an Intel Xeon Gold 6148 (2.4\,GHz), and no GPU was
used. Training used the default settings of Section~\ref{sec:tab4}.

\begin{table}[htbp]
\caption{Processing time for one validation cycle}
\label{tab:timing}
\centering
\begin{tabular}{lrr}
\hline
\textbf{Stage} & \textbf{Time (s)} & \textbf{Share} \\
\hline
Pose extraction (MediaPipe, 4 sessions) & 389.4 & 62.9\,\% \\
Joint-angle computation & 3.5 & 0.6\,\% \\
Time alignment & 0.1 & ${<}0.1$\,\% \\
Model training (50 epochs) & 225.0 & 36.4\,\% \\
Evaluation & 0.7 & 0.1\,\% \\
\hline
\textbf{Total} & 618.8 & 100\,\% \\
\hline
\end{tabular}
\end{table}

Pose extraction dominates at about 84\,ms per frame, which for 30-fps video is
roughly two and a half times the duration of the recording, and it grows linearly
with recording length. Training accounts for almost all of the remaining time. It
grows with the amount of training data and the number of epochs, and here took
under four minutes. Joint-angle computation, alignment and evaluation together
took less than five seconds. Because extraction results are cached per recording
(Section~\ref{sec:tab1}), extraction is needed only once per session. Other interactions, such as choosing different targets angles or sensor channels and retraining, cost only the training and evaluation stages.

%--------------------------------------------------------------
\subsection{Accuracy of the Ground Truth}
%--------------------------------------------------------------

The accuracy of the toolchain is bounded by the pose estimator. The original
BlazePose paper reports keypoint detection accuracy as PCK@0.2
\cite{bazarevsky2020blazepose} and does not report joint-angle error, so it cannot
serve as a reference for the accuracy of our toolchain. Lafayette et al.
\cite{lafayette2023validation} compared MediaPipe with a Qualisys system over
upper- and lower-limb movements and reported a mean absolute angular error of
$8.57\pm3.06^\circ$ over all movements ($9.98\pm3.79^\circ$ for the upper limbs and
$7.16\pm1.21^\circ$ for the lower limbs). Dill et al. \cite{dill2023mediapipe} found
that the accuracy of MediaPipe's 3D estimates depends strongly on the camera
viewing angle and on the exercise, and that it degrades quickly under less
favourable conditions.

Based on these results, we give users of the toolchain two recommendations.
Absolute joint angles obtained from a single camera in this software should not
be treated as clinical measurements. The intended use is a relative comparison
between garment revisions recorded under the same conditions. The per-channel confidence of Section~\ref{sec:tab2} also gives a relative score for showing the user how reliable each frame and each channel is.

\section{Usability Evaluation}
\label{sec:workshop}
%==============================================================

We conducted a user study to evaluate whether people can complete a full
validation workflow with the toolchain, and whether they become faster when they
repeat it.

%--------------------------------------------------------------
\subsection{Participants and Setting}
%--------------------------------------------------------------

Ten participants (P1--P10) took part. Their self-rated programming experience
ranged from 1 to 5 on the 0--5 scale described below (median 3), and three rated
themselves at 2 or below. The study was run on two laptops on which the
application had been installed and the example recordings copied in advance. No
name or other personal information was recorded.

\begin{figure*}[t]
  \centering
  \begin{minipage}[t]{0.32\textwidth}\centering
    \includegraphics[width=\linewidth]{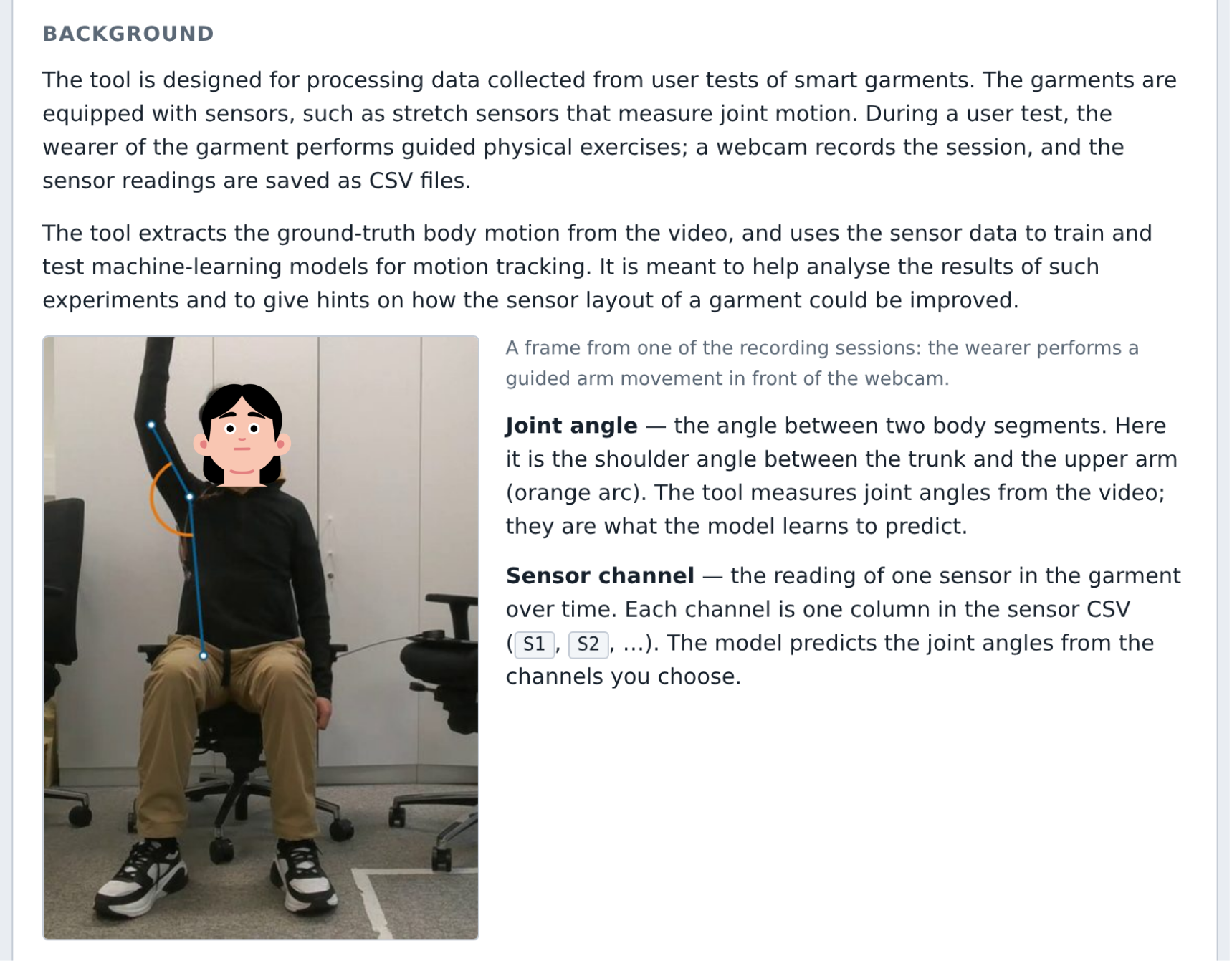}\\[1pt]
    {\footnotesize (a) Background and terms}\end{minipage}\hfill
  \begin{minipage}[t]{0.32\textwidth}\centering
    \includegraphics[width=\linewidth]{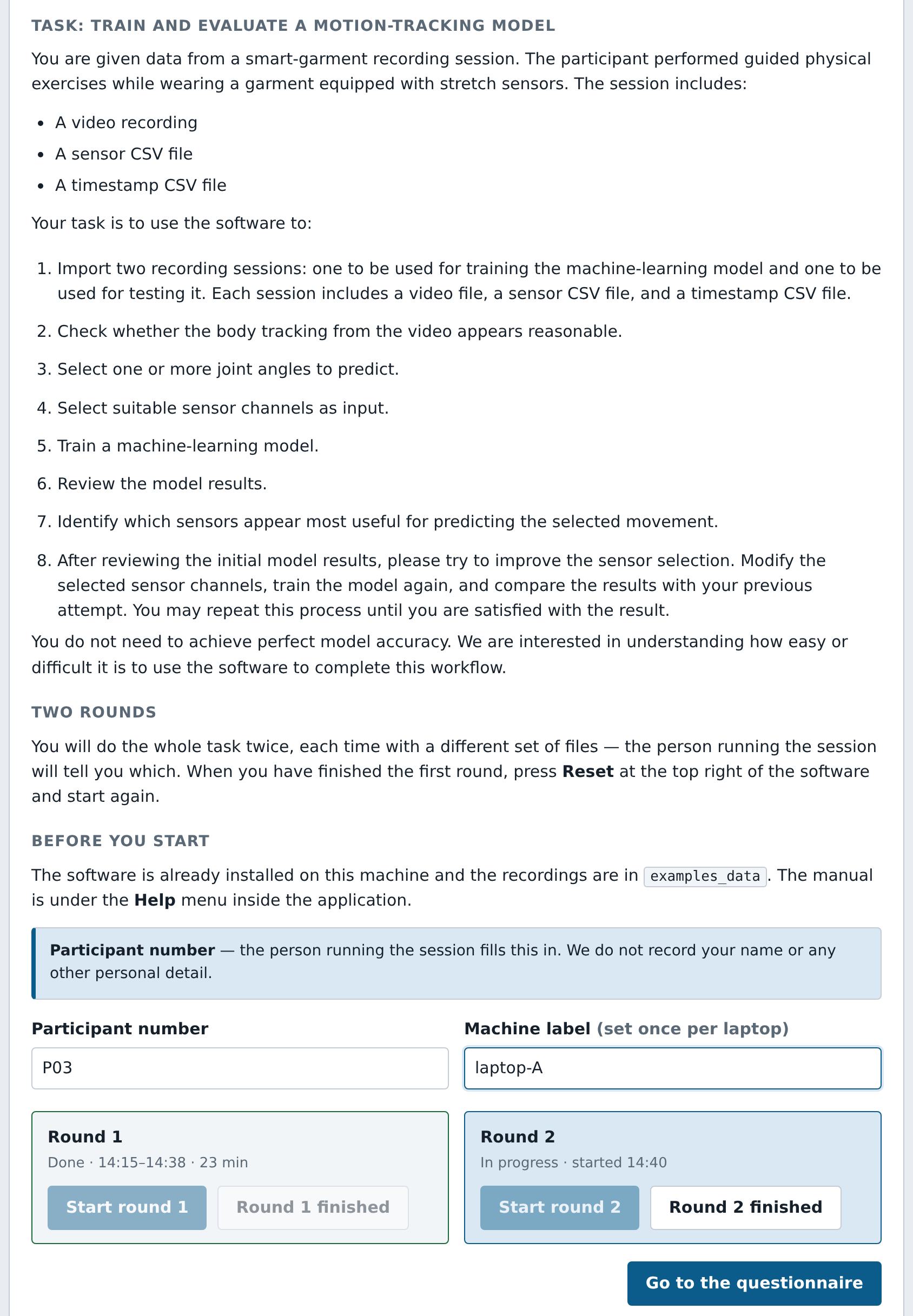}\\[1pt]
    {\footnotesize (b) Task description and round timers}\end{minipage}\hfill
  \begin{minipage}[t]{0.32\textwidth}\centering
    \includegraphics[width=\linewidth]{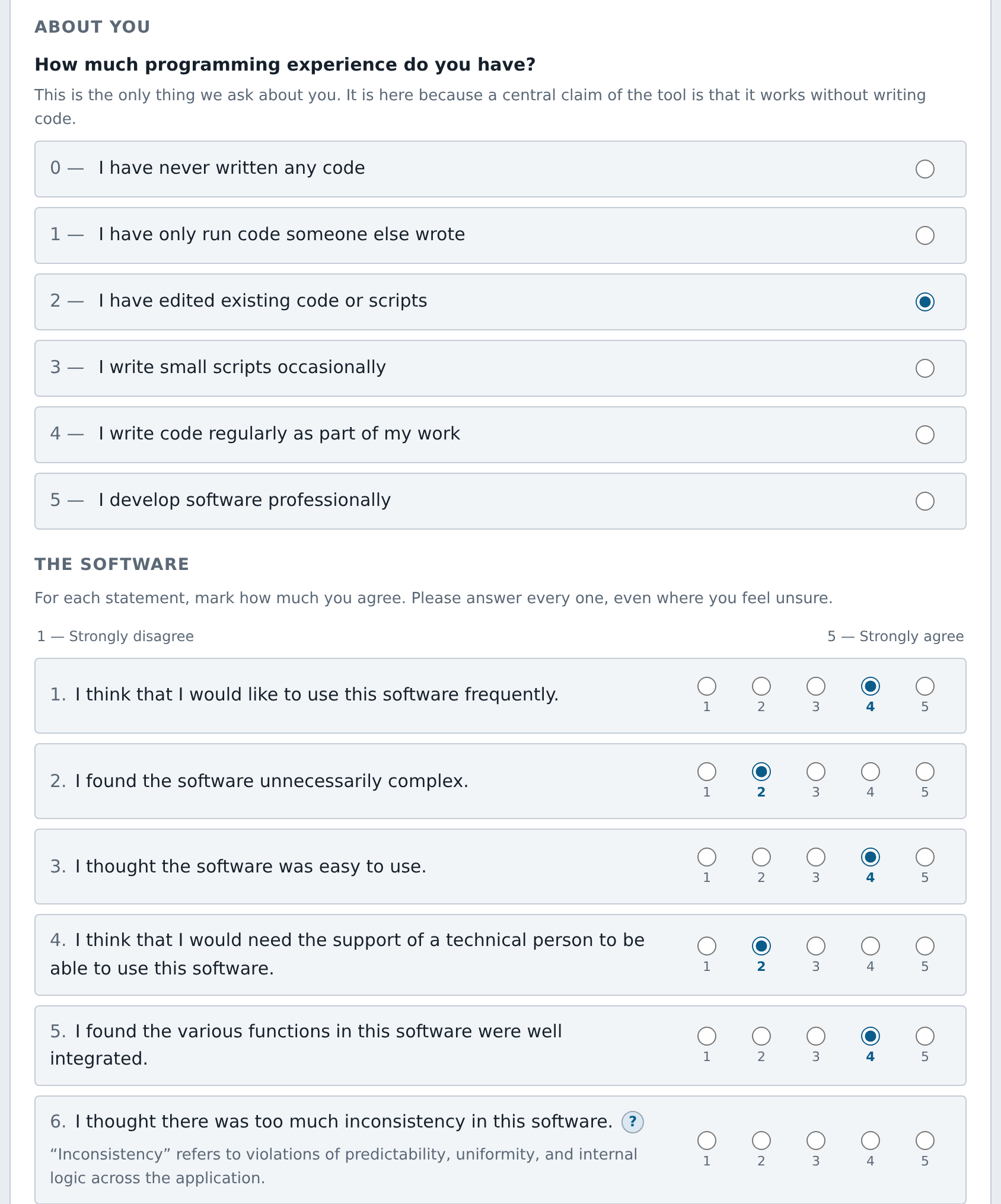}\\[1pt]
    {\footnotesize (c) Questionnaire}\end{minipage}
  \caption{The study web page. (a) The briefing introduces smart-garment user tests
  and explains, on a frame from an example recording, what a joint angle and a
  sensor channel are. (b) The task is given as a list of goals and does not name
  any tab or button. Start and end buttons record the duration of each round (the
  times shown are illustrative). (c) After both rounds, participants report their
  programming experience and complete the System Usability Scale. The explanation
  under item~6 is revealed by the question-mark button.}
  \label{fig:study}
\end{figure*}

The study was conducted as a task-based usability test with minimal guidance. The
study materials were presented on a single self-contained web page
(Fig.~\ref{fig:study}). The first part of the page briefed participants on the
context. Participants were then given the task ``Train and evaluate a
motion-tracking model'' as a list of goals: import one training and one test
session, check that the body tracking appears reasonable, select one or more
joint angles to predict and suitable sensor channels as input, train a model,
review the results, and finally try changing the sensor selection and retraining
the model. Each participant performed the task twice, since repeating the task
shows whether the workflow is learned after a single use. The duration
of each round was recorded. The built-in manual and help buttons were available throughout.

After the second round, participants completed a short questionnaire
(Fig.~\ref{fig:study}c) with three parts.
\begin{itemize}
\item Programming experience on a scale from 0 (``I have never written any
code'') to 5 (``I develop software professionally'').
\item The System Usability Scale (SUS) \cite{brooke1996sus}, ten statements rated
from 1 (strongly disagree) to 5 (strongly agree).
\item Two open questions, ``What do you like about the software?'' and ``What
aspects would you like to see improved?''
\end{itemize}

SUS scores were computed in the standard way. Odd-numbered (positively worded)
items contribute their rating minus one, even-numbered items contribute five minus
their rating, and the sum is multiplied by 2.5, giving a score from 0 to 100. We
interpret the scores against the commonly used average of 68
\cite{sauro2016quantifying} and the adjective rating scale of Bangor et al.
\cite{bangor2008sus,bangor2009adjective}. Given the small sample, we report the
median and range alongside the mean. Learnability is assessed by comparing the
durations of each participant's two rounds with a two-sided Wilcoxon signed-rank
test. We also examine whether SUS scores vary with programming experience
(Spearman's $\rho$). Answers to the open questions were grouped into themes.

%--------------------------------------------------------------
\subsection{Results}
%--------------------------------------------------------------

All ten participants finished both rounds. Table~\ref{tab:sus} lists the
results per participant and Table~\ref{tab:rounds} summarises the two rounds.

\begin{table}[htbp]
\caption{Results per participant}
\label{tab:sus}
\centering
\begin{tabular}{lcccc}
\hline
\textbf{P} & \textbf{Coding} & \textbf{Round 1} & \textbf{Round 2} & \textbf{SUS} \\
 & (0--5) & (min) & (min) & (0--100) \\
\hline
P1  & 3 & 8.5 & 3.0 & 32.5 \\
P2  & 4 & 8.3 & 2.9 & 60.0 \\
P3  & 3 & 7.4 & 2.1 & 70.0 \\
P4  & 2 & 8.1 & 2.5 & 67.5 \\
P5  & 5 & 6.8 & 1.9 & 72.5 \\
P6  & 4 & 7.6 & 2.3 & 60.0 \\
P7  & 1 & 8.7 & 2.8 & 70.0 \\
P8  & 3 & 6.5 & 1.8 & 75.0 \\
P9  & 4 & 7.9 & 2.4 & 67.5 \\
P10 & 2 & 7.2 & 2.0 & 72.5 \\
\hline
\textbf{Median} & 3 & 7.8 & 2.3 & 68.8 \\
\textbf{Range} & 1--5 & 6.5--8.7 & 1.8--3.0 & 32.5--75.0 \\
\hline
\end{tabular}
\end{table}

\begin{table}[htbp]
\caption{Duration of the first and second round ($N=10$)}
\label{tab:rounds}
\centering
\begin{tabular}{lcc}
\hline
\textbf{Measure} & \textbf{Round 1} & \textbf{Round 2} \\
\hline
Median duration (min)  & 7.8 & 2.3 \\
Range (min)            & 6.5--8.7 & 1.8--3.0 \\
Mean duration (min)    & 7.7 & 2.4 \\
\hline
Reduction per participant & \multicolumn{2}{c}{median 70\,\% (range 65--72\,\%)} \\
Wilcoxon signed-rank   & \multicolumn{2}{c}{$W=0$, $p=0.002$} \\
\hline
\end{tabular}
\end{table}

Every participant completed the second round faster than the first. The median
duration fell from 7.8 to 2.3~minutes, and the individual reductions were
consistent, between 65 and 72\,\% ($W=0$, $p=0.002$). After a single pass,
participants could repeat the whole workflow in about two and a half minutes,
including importing two sessions, selecting targets and inputs, training the
model and reading the results.

The mean SUS score was 64.8 (SD 12.4, 95\,\% CI 55.9--73.6) and the median 68.8,
with five of the ten participants at or above the average of 68
\cite{sauro2016quantifying}. The mean therefore lies slightly below that benchmark
and the median at it, between the ``OK'' and ``good'' adjective ratings of Bangor
et al. \cite{bangor2009adjective}. The spread comes mainly from one participant.
P1 rated the software at 32.5, while the other nine scores lie between 60 and 75
(median 70.0). At the item level, the strongest agreement was with item~7 (``most
people would learn to use this software very quickly'', mean 3.9 on the 1--5
scale) and item~5 (``functions well integrated'', 3.7). The strongest
disagreement was with item~4 (``I would need the support of a technical person'',
2.1) and item~2 (``unnecessarily complex'', 2.1). Among the negatively worded
items, item~10 (``I needed to learn a lot of things before I could get going'',
3.3) received the highest agreement, which is consistent with a first round that
took three times as long as the second.

SUS scores showed no relationship with programming experience (Spearman's
$\rho=-0.19$, $p=0.60$). The three participants who rated their experience at 2
or below scored 67.5, 70.0 and 72.5 (median 70.0), compared with a median of 67.5
for the seven with more experience. Ten participants are too few to show that
experience makes no difference, but the results give no indication that the tool
is harder to use for people who program less.

Five participants answered at least one of the open questions. They liked the absence of code (``no coding, all of the operations are GUI''), the division of the workflow into steps ``that can be changed anytime'', and a ``well-structured layout''. Their suggestions fell into two recurring themes. The first is interpretation. Two participants asked the tool to explain ``what the results mean and what I can do with it'', or to provide ``a more in-depth analysis of the results'' for designers and other non-technical users. The second is a more detailed introduction for the toolchain.
Two participants asked for guidance on first use, either as a tutorial inside the application or as ``a brief video or demo''. Two further suggestions concerned individual features, namely showing the skeleton of the second tab side by side with the sensor readings of the third, and an undo function in addition to Reset.
The software is still under active development. We will keep releasing new versions at GitHub to address these requests, and its issue tracker is open to the needs of future users.

%--------------------------------------------------------------
\subsection{Limitations}
%--------------------------------------------------------------

Our user study also has several limitations. Because we do not collect any personally identifiable information and recruiting participants proved challenging, not all participants are professional garment designers. The second round always followed the first, as the learnability question requires, so it also benefited from general familiarity with the task. The application was pre-installed, so the study does not measure the effort of installing the bundle. All recordings came from one wearer and one garment, and they were prepared in advance, so the study covers the analysis workflow and leaves data collection untested.

%==============================================================
\section{Conclusion and Future Work}
\label{sec:conclusion}
%==============================================================

We presented a no-code toolchain that enables non-technical e-textiles
researchers and practitioners to validate wearable motion sensing systems with a
simple, low-cost setup consisting of a standard laptop and a webcam. The
toolchain integrates webcam-based ground-truth extraction, machine learning model
training, and performance analysis and visualisation into a single graphical
workflow, allowing designers to iterate on sensor selection and layout choices
without motion-capture infrastructure or programming expertise. We also conducted
a user study to evaluate the usability of the toolchain and to identify
directions for further improvement.

Several directions remain for future work. First, the usability study points to
where the interface should go next. Participants could operate the workflow but
asked for help in interpreting its results, so we plan to add plain-language
explanations of what each result implies for the garment and what to try next.
Using large language models to analyse the results and suggest next steps is also a
promising direction for this. Second, the current pipeline relies on
single-camera pose estimation, which limits accuracy for out-of-plane motions and
self-occluded joints. Supporting optional multi-camera input while preserving the
simple setup is a natural extension. Third, we intend to expand the library of
built-in regression models and to extend the toolchain to classification tasks, and support a wider variety of sensor types,
which would further broaden its practical use.

%==============================================================
\section*{Acknowledgment}
%==============================================================
% TODO: funding sources, compute resources and contributors.

%==============================================================
\section*{Generative AI Declaration}
%==============================================================
Generative AI tools were used to assist in developing the software described in
this paper and to polish the language of the manuscript. The authors reviewed
and tested the software, verified the content of the manuscript, and take full
responsibility for both.

%==============================================================
% References: references.bib, formatted by the official IEEEtran.bst (v1.14,
% from CTAN). It lists entries in order of first citation, as IEEE requires.
%==============================================================
\bibliographystyle{IEEEtran}
\bibliography{references}

\end{document}